\documentclass[aps,twocolumn,a4paper,10pt]{revtex4-1}
\usepackage{mathrsfs}
\usepackage{graphicx}
\usepackage{latexsym}
\usepackage{amsmath}
\usepackage{amssymb}
\usepackage{textcomp}
\usepackage{amsbsy}
\usepackage{epstopdf}

\begin{document}
 \title{\large \bf Modulus stabilisation in a backreacted warped geometry model via Goldberger-Wise mechanism}
\author{Ashmita Das}
\email{ashmita@iitg.ernet.in}
\affiliation{Department of Physics, Indian Institute of Technology, North Guwahati, Guwahati, Assam 781039, India}
\author{Tanmoy Paul}
\email{tpap@iacs.res.in}
\author{Soumitra SenGupta}
\email{tpssg@iacs.res.in}
\affiliation{Department of Theoretical Physics,\\
Indian Association for the Cultivation of Science,\\
2A $\&$ 2B Raja S.C. Mullick Road,\\
Kolkata - 700 032, India.\\}

\begin{abstract}
In the context of higher dimensional braneworld scenario, 
the stabilisation of extra dimensional modulus  is an
essential requirement for resolving the gauge hierarchy problem in the context of Standard Model of elementary particle Physics. 
For Randall-Sundrum (RS) warped extra dimensional model, Goldberger 
and Wise (GW) proposed a much useful mechanism to achieve this using a scalar field 
in the bulk spacetime ignoring the effects of backreaction of the scalar field on the background metric.
In this article we examine the influence of the backreaction of the stabilising field on the stabilisation
 condition as well as that on the  Physics of the extra dimensional modulus namely radion.  In particular 
we  obtain the modifications of the 
 mass and the coupling of the radion with the Standard Model (SM) matter fields
on the TeV brane due to backreaction effect. Our calculation also brings out
an important equivalence between the treatments followed by Csaki et.al. in
 \cite{kribs} and Goldberger-Wise in \cite{GW1,GW_radion}.
\end{abstract}
\maketitle

\section*{Introduction}
The gauge hierarchy problem in SM of particle Physics results into the well known fine
tuning problem in connection to the Higgs mass which acquires a quadratic divergence due to
the large radiative correction in perturbation theory.
In order to confined Higgs mass parameter within TeV scale, one needs to consider theories beyond the 
SM of particle Physics.
Among many such attempts \cite{arkani,horava,RS1,kaloper,cohen,burgess,chodos}, Randall-Sundrm (RS) 
warped extra dimensional scenario
has earned special attention for 
following reasons \cite{RS1}:
\begin{itemize}
 \item
 It resolves the gauge hierarchy problem 
without introducing any other intermediate scale in the theory.
\item
The modulus of the extra dimension can be stabilised by introducing
 a bulk scalar field\cite{GW1}.
 \end{itemize}
Warped solutions, similar to RS model, can also be found
from string theory which predicts inevitable existence of extra dimensions \cite{Green}.

In search for such extra dimensions, the detectors in LHC \cite{atlas1,atlas2}
 have been designed to explore possible signatures of the warped
geometry models through phenomenology of  RS graviton \cite{dhr,rizzo,yong,dhr1,thomas}, 
radion \cite{GW_radion,kribs,julien,wolfe} and  RS black holes \cite{wiseman,naresh,dai}.\\
One of the crucial aspect of this braneworld model is to stabilise the distance between the two
branes ( known as radius modulus or radion ). For this one needs to generate an appropriate 
potential term for the radion field with a stable minimum consistent with the value  proposed in RS model
in order to solve the gauge hierarchy problem.
Goldberger and Wise proposed a very useful mechanism to achieve this by 
introducing a bulk scalar field with appropriate boundary terms. They showed that one can indeed
stabilise the modulus without any unnatural fine tuning of the parameters when the effects of the backreaction 
of the bulk scalar  on the background metric can be ignored. Subsequently the phenomenology of the
resulting radion field as a fluctuation about this stable minimum of the modulus became an important
area of study specially in the context of collider phenomenology of extra dimensional 
scenario beyond Standard Model of particle Physics.\\
The important questions that however remain are, 
\begin{itemize}
\item when the energy momentum tensor of the bulk scalar
is significantly large so that it's backreaction  on the bulk 5-dimensional metric can not be ignored, 
can we still stabilise the modulus following GW prescription ? What is the resulting stabilisation 
condition?
\item If the modulus is stabilised, how does the  mass and couplings of the radion field change 
due to the  backreaction effect?
\end{itemize}
We aim to address these questions in this work. After a brief review of RS model,
we obtain the modulus potential generated by integrating  the bulk scalar field action in a fully 
backreacted 5-dimensional metric. We explicitly find the expression for the radion minimum which matches 
with that proposed in \cite{kribs} where our result additionally determines the exact form of the boundary 
value of the scalar field at the TeV brane.
We also  obtain the radion mass and coupling with standard model field following the procedure 
proposed in \cite{GW_radion}. We conclude by comparing our results with that obtained in \cite{kribs}.

\section*{Brief description of RS scenario and its stabilisation via GW mechanism}
RS scenario is defined on a five dimensional spacetime involving one warped and compact 
extra spacelike dimension. This scenario postulates gravity in the five-dimensional
`bulk', whereas our four-dimensional universe is confined to one of the two 3-branes
known as TeV/visible and Planck/hidden brane 
located at the two orbifold fixed points along the compact dimension.\\
The RS model is characterised by the non-factorisable 
background metric,
\begin{equation}
ds^2 = e^{- 2 kr_c|\phi|} \eta_{\mu\nu} dx^{\mu} dx^{\nu} -r_c^2d\phi^2 \label{eq1}
\end{equation}
 The extra dimensional angular coordinate is denoted by  $\phi$ and  ranges from
$-\pi$ to $+\pi$ following a $S^1/Z_2$ orbifolding.
Here, $r_c$ is the compactification radius of the extra dimension. Two 3-branes are 
located at the orbifold fixed points $\phi=(0,\pi)$. 
The quantity $k=\sqrt{\frac{-\Lambda}{12M^3}}$, which is of the order of 5-dimensional Planck
scale $M$. Thus $k$ relates the 5D Planck scale $M$ to the 5D cosmological constant
$\Lambda$.\\
The hidden and visible brane tensions are,
$V_{hid}=-V_{vis}=12M^3k$.
All the dimensionful parameters described above are
related to the reduced 4-dimensional Planck scale ${M}_{Pl}$ as,
\begin{equation}
 M_{Pl}^2=\frac{M^3}{k}(1-e^{-2k\pi r_c})\label{rplanckmass}
\end{equation}
For $k\pi r_c \approx 36$, the exponential factor present
in the background metric, which is often called warp factor, produces 
a large suppression so that a mass scale of the order of Planck scale is reduced to TeV scale on the 
visible brane. 
A scalar mass say mass of Higgs is given as, 
\begin{equation}
 m_H=m_{0}e^{-k\pi r_c}\label{physmass}
\end{equation}
Here, $m_H$ is Higgs mass parameter 
on the visible brane and $m_0$ is the natural scale of the theory
above which new physics beyond SM is expected to appear \cite{RS1}. \\ 

In higher dimensional braneworld scenario, the stabilisation of extra dimensional
modulus is a crucial aspect and needs to be addressed carefully. 
It has been demonstrated by Goldberger and Wise that the modulus corresponding to the
 radius of the extra dimension in RS warped geometry model can be stabilised \cite{GW1} by invoking a massive 
scalar field in the bulk. Consequently the phenomenology of the radion 
field originating from 5D gravitational degrees of freedom has also been explored \cite{GW_radion}.
GW mechanism for stabilisation postulates a bulk scalar field with different vacuum expectation values
(VEV) at the two 3-branes that reside at the orbifold fixed points of
$S_1/Z_2$ compactification. This mechanism, however, generates a bulk energy
density which may modify the warped geometry itself via backreaction. Though 
this had been neglected in the initial GW proposal, its various implications,
including the modifications of the warp factor, 
have been subsequently investigated \cite{kribs,wolfe}.
The authors of \cite{kribs} have explored the radion phenomenology in the 
background of the backreacted RS warped geometry model assuming the boundary values
of the scalar field at the two branes.  They
proposed that the extra dimensional modulus will be stabilised at
 $k\pi r_0=\frac{k}{u}{\rm ln}(\frac{\Phi_P}{\Phi_T})$ where $r_0$ is the stabilised value of the modulus,
 $u$ is the quartic coupling parameter in the bulk scalar field potential and $\Phi_P$, $\Phi_T$ are the boundary values of 
 the scalar field at the two orbifold fixed points where the Planck and TeV branes are located.\\
In this work, we want to explore whether the modulus stabilisation of such backreacted warped scenario 
can also be analysed  via GW mechanism with a quartic potential for the 
stabilising scalar field in the bulk as described in \cite{kribs}.
Considering the quartic form of potential for the bulk stabilising scalar field,  
we further explore the radion phenomenology in this background model by using the technique
as explained in \cite{GW_radion}. We then compare our results with \cite{GW1,GW_radion, kribs}. 
We organise our work as follows: In section {\ref{BR_RS}}, 
we describe five dimensional warped geometry model which includes the effect of the  
backreaction of the bulk stabilising scalar field on the background geometry.
In section {\ref{stabilisation_BR}}, we explain the stabilising mechanism
of this backreacted warped geometry model by using the GW mechanism.
In section {\ref{radion_mass}, \ref{radion_coupling}}, we find the radion mass and its coupling
with the SM matter fields on TeV brane which bring out the modifications of these due to back-reaction effect.
\section{Backreacted Randall-Sundrum model}\label{BR_RS}
We consider the action for this background geometry as,
\begin{eqnarray}
 S&=&-M^3\int d^5x \sqrt{G}[R-\Lambda]\nonumber\\
&+&\int d^5x \sqrt{G}[(1/2)G^{MN}\partial_M\Phi\partial_N\Phi-V(\Phi)]\nonumber\\
 &-&\int d^4x \sqrt{-g_{hid}}\lambda_{hid}(\Phi) - \int d^4x \sqrt{-g_{vis}}\lambda_{vis}(\Phi)
 \label{action}
\end{eqnarray}
where $M$ is the five dimensional Planck scale, $G_{MN}$ is the 
five dimensional metric where $g_{hid}$ and $g_{vis}$ are the induced 
metric on hidden and visible brane respectively. $\Lambda$ 
symbolises the bulk cosmological constant, $\Phi$ is the 
scalar field and $V(\Phi)$ is the scalar field potential. 
 $\lambda_{hid}$, $\lambda_{vis}$  are the self interactions of scalar field (including brane 
tensions) on Planck, TeV branes. 
We consider the background metric ansatz as,
\begin{equation}
 ds^2 = \exp{[-2A(\phi)]}\eta_{\mu\nu}dx^{\mu}dx^{\nu} - r_c^2d\phi^2
 \label{metric ansatz}
\end{equation}
where $A(\phi)$ is the warp factor. For simplicity we assume that 
the bulk scalar field depends only on the extra dimensional coordinate ($\phi$). 
Thus the 5-dimensional Einstein's and scalar field equations 
for this metric can be written as,
\begin{eqnarray}
 \frac{4}{r_c^2} A'^2(\phi) - \frac{1}{r_c^2}A''(\phi)&=&-(2\kappa^2/3)V(\Phi)\nonumber\\ 
&-&(\kappa^2/3)\sum \lambda_{i}(\Phi)\delta(\phi-\phi_{i})
 \label{equation1}
\end{eqnarray}

\begin{equation}
  \frac{1}{r_c^2}A'^2(\phi)= \frac{\kappa^2}{12r_c^2}\Phi'^2 -(\kappa^2/6)V(\Phi)
 \label{equation2}
\end{equation}

\begin{eqnarray}
 \frac{1}{r_c^2}\Phi''(\phi)=\frac{4}{r_c^2}A'\Phi' + \frac{\partial V}{\partial \Phi} 
+ \sum \frac{\partial \lambda_{i}}{\partial \Phi}\delta(\phi-\phi_{i})
 \label{equation3}
\end{eqnarray}
Where  $M^3=(1/2\kappa^2)$. Here index $i$ is used to designate the two branes and prime denotes the derivative with 
respect to $\phi$. From the above equations, 
the boundary conditions of $A(\phi)$ and $\Phi(\phi)$ are obtained as,
\begin{equation}
 \frac{1}{r_c}[A'(\phi)]_{i} = (\kappa^2/3)\lambda_{i}(\Phi_{i})
 \label{bc1}
 \end{equation}
 and
 \begin{equation}
  \frac{1}{r_c}[\Phi'(\phi)]_{i}=\partial_{\Phi}\lambda_{i}(\Phi_{i})
 \label{bc2}
 \end{equation}
Square bracket in the above two equations represents the jump of the
 variables $A'(\phi)$ and $\Phi'(\phi)$ at the branes. In order to get an analytic 
solution of backreacted Randall-Sundrum scenario, let us consider the form of the scalar field potential 
as \cite{kribs},
\begin{equation}
 V(\Phi) = (1/2)\Phi^2(u^2+4uk) - (\kappa^2/6)u^2\Phi^4
 \label{scalar potential}
\end{equation}
where $k=\sqrt{-\kappa^2\Lambda/6}$. The potential contains quadratic
 as well as quartic self interaction of the scalar field. Moreover it may be 
noticed that the mass and quartic 
coupling of the field $\Phi(\phi)$ are connected by a common free parameter $u$. 
Using this form of the potential, one can obtain a 
solution of $A(\phi)$ and $\Phi(\phi)$ as follows,
\begin{eqnarray}
 A(\phi) = kr_c|\phi| + (\kappa^2/12)\Phi_{P}^2\exp{(-2ur_c|\phi|)}
 \label{solution of warp factor}\\
 \Phi(\phi) = \Phi_{P}\exp{(-ur_c|\phi|)}
 \label{solution of scalar field}
\end{eqnarray}
where $\Phi_{P}$ is taken as the value of the scalar field on the Planck brane.
 Moreover $\lambda_{hid}$ and $\lambda_{vis}$ can be obtained from the 
boundary conditions (eqn.(\ref{bc1}) and eqn.(\ref{bc2})) as,
\begin{eqnarray}
 \lambda_{hid} = 6k/\kappa^2 - u\Phi_{P}^2
 \label{planck brane tension}\\
 \lambda_{vis} = -6k/\kappa^2 + u\Phi_{P}^2\exp{(-2u\pi r_c)}
 \label{planck brane tension}
 \end{eqnarray}
 
 \section{Stabilisation mechanism}\label{stabilisation_BR}
We now address the modulus stabilisation using GW prescription by including the 
effects of the backreaction of the stabilising bulk scalar on the background geometry.
We aim to determine how the backreaction can influence the stability of the braneworld.\\
 Plugging the scalar field solution from eq.(\ref{solution of scalar field}) 
into the five dimensional scalar field action 
and integrating over $\phi$, we obtain an effective four dimensional potential for 
 $r_c$ as,
 \begin{eqnarray}
  V_{eff}(r_c)&=&r_c\int d\phi \exp{[-4A(\phi)]} [u^2\Phi_{P}^2\exp{(-2ur_c\phi)}\nonumber\\
&+&(u^2+4uk)\Phi_{P}^2]\exp{(-2ur_c\phi)}- (\kappa^2u^2/3)\Phi_{P}^4]\nonumber\\
&\exp&{(-4ur_c\phi)}]+ \exp{[-4A(0)]}u\Phi_{P}^2\nonumber\\
&-&\exp{[-4A(\pi r_c)]}u\Phi_{P}^2\exp{(-2u\pi r_c)}
  \label{effective potential}
  \end{eqnarray}
The stabilised value for the inter brane separation can be achieved
by minimising the low energy effective potential for the modulus field $r_c$.
The minimisation of the low energy effective potential with respect to $r_c$
 can be obtained (by using Leibniz's theorem) as follows,
  \begin{eqnarray}
   \frac{\partial V_{eff}}{\partial(\pi r_c)}&=&\exp{[-4A(\pi r_c)]}\exp{(-2u\pi r_c)}\\
   &[&4u^2\Phi_{P}^2+8uk\Phi_{P}^2-\kappa^2u^2\Phi_{P}^4\exp{(-2u\pi r_c)}]
   \nonumber\\
  \end{eqnarray}
  From the above expression we obtain the stabilisation condition for the modulus field
$r_c$ as,
  \begin{equation}
   k\pi r_c = \frac{k}{u}\ln{\{\frac{\kappa\Phi_{P}}{2\sqrt{1+\frac{2k}{u}}}\}}
   \label{stabilized modulus}
  \end{equation}
  Equation(\ref{stabilized modulus}) implies the stabilisation condition
between the two 3-branes in the backreacted Randall-Sundrum set up. Comparing the scalar field solution 
  (eqn.(\ref{solution of scalar field})) and the expression of stabilised
 modulus (eqn.(\ref{stabilized modulus})), the expression for  
 the VEV of scalar field on TeV brane ({\it {i.e}} $\Phi_{T}$)
can be written as $\kappa\Phi_{T} = 2\sqrt{1+\frac{2k}{u}}$. 
With this  value of the scalar field on the TeV brane , eqn.(\ref{stabilized modulus}) becomes 
$k \pi r_c = \frac{k}{u} \ln(\Phi_{P}/\Phi_T)$ which matches with that described in \cite{kribs}.\\
Our entire analysis of finding the stabilisation condition in eqn.(\ref{stabilized modulus}) 
is valid only for $u>0$. In this context one can easily 
check that $V_{eff}(r_c)$ produces no minima for $u<0$. Hence the parameter 
  $u$ is confined in positive regime in order to make a stable configuration for this
backreacted braneworld scenario.

As we have seen earlier that in order to stabilise RS braneworld scenario, GW in \cite{GW1}
has chosen a quadratic potential for the bulk stabilising scalar field.
However the backreacted model that we have considered contains also quartic term 
of the bulk stabilising scalar field potential.
If we keep only the leading order terms in $u$, the potential in eqn.(\ref{scalar potential}) tends 
to a quadratic potential where the mass of the
 scalar field ($\Phi$) can be written as $m_{\Phi}^2=4uk$. In this limit 
 the stabilisation condition becomes,
  \begin{equation}
   k\pi r_c = \frac{4k^2}{m_{\Phi}^2}\ln{\{\frac{\kappa\Phi_{P}}{2\sqrt{1+\frac{2k}{u}}}\}}
   \nonumber\\
  \end{equation}
  which is same as the expression for the stabilised modulus obtained in \cite{GW1}. 
For $u/k$ less than unity, we neglect the higher orders of $u/k$ which in turn 
implies that the  scalar potential contains 
  only the mass term ($V(\Phi)=2uk\Phi^2$ just as GW original work of stabilisation 
  ( see  eqn.(\ref{scalar potential})). 
Furthermore, for small $u/k$, the ratio between the energy momentum tensor of 
  the scalar field and of the bulk cosmological constant $\Lambda$ 
goes as $u/k$ which immediately justifies the assumptions made by Goldberger and Wise 
that the scalar field backreaction is negligible with respect to 
  $\Lambda$. This explains why our final result of stabilised inter brane 
  separation (eqn.(\ref{stabilized modulus})) converges exactly to the stabilised modulus 
expression obtained in \cite{GW1} in the limit of small $u/k$ parameter. However we 
emphasise that the 
expression in eqn.(\ref{stabilized modulus}) is a  generalised stabilisation condition 
including the  full backreaction of the stabilising field.
\section{Radion mass}\label{radion_mass}
  In this section, we consider a small fluctuation of the  brane locations 
 around the stable inter brane separation $r_c$ which
depends on the brane coordinates $x^{\mu}$. \\
The corresponding metric ansatz is,
  \begin{equation}
   ds^2 = \exp{[-2A(x,\phi)]}g_{\mu\nu}(x)dx^{\mu}dx^{\nu} - T^2(x)d\phi^2
   \label{metric ansatz 2}
  \end{equation}
  where $T(x)$ measures the inter brane fluctuations between the branes  and $\phi$ is 
the extra dimensional angular coordinate.\\
The modified warp factor therefore can be written as, 
  \begin{equation}
   A(x,\phi) = k|\phi|T(x) + \frac{\kappa^2\Phi_{P}^2}{12}\exp{[-2u|\phi|T(x)]}
   \label{warp factor 2}
  \end{equation}
 In the four dimensional effective theory, $T(x)$ appears as an additional scalar field 
 known as modulus field \cite{GW_radion}. A Kaluza-Klein reduction of the five dimensional 
  Einstein-Hilbert action for the above warp factor (eqn.(\ref{warp factor 2})) 
leads to the kinetic term of $T(x)$ as follows:
  \begin{eqnarray}
   &S_{kin}[T]=12M^3 \int d^4x \sqrt{-g}\int d\phi \exp{[-2A(x,\phi)]}\nonumber\\
   & [k\phi\partial_\mu T\partial^\mu T
   (1-\frac{\kappa^2\Phi_{P}^2}{6}\frac{u}{k}\exp{(-2u\phi T)})\nonumber\\
   & - k^2\phi^2T\partial_\mu T\partial^\mu T(1-\frac{\kappa^2\Phi_{P}^2}{6}\frac{u}{k}\exp{(-2u\phi T)})^2]
   \label{kinetic part of radion 1}
  \end{eqnarray} 
  Assuming that $\frac{\kappa^2\Phi_{P}^2}{6}\frac{u}{k}$ is
less than unity such that $(1-\frac{\kappa^2\Phi_{P}^2}{6}\frac{u}{k}) \simeq 1$, 
and integrating over the extra dimensional coordinate ($\phi$),\\
we obtain,
  \begin{eqnarray}
   S_{kin}[T]&=&\frac{6M^3}{k} \int d^4x \sqrt{-g}\partial_{\mu}(\exp{[-A(\pi,x)]})\nonumber\\
&\partial^{\mu}&(\exp{[-A(\pi,x)]})
   \label{kinetic part of radion 2}
  \end{eqnarray}
In order to obtain a canonically normalised radion field, we now redefine $T(x) \longrightarrow \psi(x)$ where
  \begin{equation}
   \Psi(x) = \sqrt{\frac{12M^3}{k}}\exp{[-A(x,\pi)]}
   \label{transformation of field}
  \end{equation}
  With respect to this redefined field ($\Psi(x)$), the kinetic part of the action becomes canonical as,
  \begin{equation}
   S_{kin}[\Psi] = (1/2)\int d^4x \sqrt{-g} \partial_{\mu}\Psi\partial^{\mu}\Psi
   \nonumber\\
  \end{equation}
Now we turn our focus to find the radion mass square ($m_{\Psi}^2$) from the following expression,
  \begin{equation}
   m_{\Psi}^2 = [V_{eff}''(T)*T'(\Psi)^2]_{(<T>=r_c)}
   \label{radion mass 1}
  \end{equation}
  where $r_c$ is the stabilized modulus (eqn.(\ref{stabilized modulus})) 
and $V_{eff}(T)$ is obtained from eqn.(\ref{effective potential}) by 
  replacing $r_c$ by $T(x)$. Therefore, we 
  determine $V_{eff}''(T)$ and $T'(\Psi)^2$ at $<T>=r_c$ as follows,
  \begin{equation}
   V_{eff}''(r_c) = 2\kappa^2u^3\Phi_{P}^4 \exp{[-4A(\pi r_c)]}\exp{(-4u\pi r_c)}\nonumber\\
\end{equation}
\begin{equation}
   T'(\Psi)^2|_{r_c} = \frac{1}{12M^3k} \exp{[2A(\pi r_c)]}
   \nonumber\\
  \end{equation}
  where we use the assumption ($\frac{\kappa^2\Phi_{P}^2}{6}\frac{u}{k}<1$) and 
  $A(\pi r_c) = k\pi r_c+\frac{\kappa^2\Phi_{P}^2}{12}\exp{(-2u\pi r_c)}$. 
Putting these  expressions into eqn.(\ref{radion mass 1}) we
obtain the radion mass square as (taking $\frac{\kappa\Phi_{P}}{\sqrt{2}}=l$)
  \begin{eqnarray}
   m_{\Psi}^2&=&\{\frac{8}{3k} u^2l^2(2k+u)\exp{[-2(u+k)\pi r_c]}\}\nonumber\\
&\{&\exp{[-\frac{l^2}{3}\exp{(-2u\pi r_c)}]}\}
   \label{radion mass 2}
  \end{eqnarray}
Thus, we obtain the mass of the radion field ( following the procedure 
described in \cite{GW_radion} ) when the bulk scalar potential ($V(\Phi)$) 
  contains quadratic as well as quartic self interaction of the scalar 
field ($\Phi$). Our final result of radion mass (eqn.(\ref{radion mass 2})) 
 though  resembles to the mass expression in \cite{kribs} (eqn.(6.6)) however has 
an additional correction. This small correction arises because our 
  assumption ($l^2 \frac{u}{k}<1$) is slightly different 
from ($l^2<1$) which is considered by the authors of 
\cite{kribs}. However, for $l^2<1$, our result exactly
coincides with the radion mass expression given in \cite{kribs}.
  
  Again, in order to find a correlation between these two
different mechanism of finding radion mass namely in \cite{GW_radion} and in \cite{ kribs},
we determine the mass of the radion in the leading order of parameter $u/k$.
As a a result, eqn.(\ref{radion mass 2}) turns out to be,
  \begin{equation}
   m_{\Psi}^2 = \frac{4k^2\Phi_{P}^2}{3M^3} \epsilon^2\exp{(-2k\pi r_c)}
   \nonumber\\
  \end{equation}
  where $\epsilon = m_{\Phi}^2/4k^2$. The above expression is same as the radion mass obtained in \cite{GW_radion}.
 Earlier we showed that the stable value of the modulus also matched with \cite{GW1}
 in the leading order of the parameter $u/k$
  
  \section{Coupling between radion and Standard Model fields}\label{radion_coupling}
The radion field arises as a scalar degree
of freedom on the TeV brane and has interactions  with the Standard Model (SM) 
fields. 
  From the five dimensional metric ansatz (eqn.(\ref{metric ansatz 2})), 
it is clear that the induced metric on visible brane is 
  $(\frac{\Psi}{f})^2g_{\mu\nu}$ (where $f=\sqrt{12M^3/k}$) 
and consequently $\Psi(x)$ couples directly with SM fields. \\
For example, consider the 
  Higgs sector of Standard Model,
  \begin{eqnarray}
   S_{Higgs}&=&(1/2) \int d^4x \sqrt{-g}(\Psi/f)^4\nonumber\\ 
&[&(\Psi/f)^{-2}g^{\mu\nu}\partial_\mu h\partial_\nu h - \mu_0^2h^2]
   \nonumber\\
  \end{eqnarray}
  where $h(x)$ is the Higgs field. In order to get a canonical
 kinetic term, one needs to redefine $h(x) \longrightarrow H(x) = \frac{<\Psi>}{f}h(x)$. 
Therefore for $H(x)$, the above action can be written as,
  \begin{eqnarray}
   S_{Higgs}&=&(1/2) \int d^4x \sqrt{-g} [(\frac{\Psi}{<\Psi>})^{2}g^{\mu\nu}\partial_\mu H\partial_\nu H\nonumber\\
&-&(\frac{\Psi}{<\Psi>})^{4}\mu^2H^2]
   \label{higgs action}
  \end{eqnarray}
  where $\mu = \mu_0 \frac{<\Psi>}{f} = \mu_0\exp{[-A(\pi r_c)]}$. 
Considering a fluctuation of $\Psi(x)$ about its VEV as 
  $\Psi(x)=<\Psi> + \delta\Psi$, one can obtain (from eqn.(\ref{higgs action})) 
that $\delta\Psi$ couples to $H(x)$ through the trace of the 
  energy-momentum tensor of the Higgs field:
  \begin{equation}
   \mathcal{L} = \frac{\delta\Psi}{\Psi}T^\mu_\mu(H)
   \nonumber\\
  \end{equation}
  So, the coupling between radion and Higgs field become, 
 $\lambda_{(H-\delta\Psi)} = \frac{\mu^2}{<\Psi>}$. Similar consideration holds for any other 
  SM fields. For example for $Z$ boson, the corresponding coupling is 
$\lambda_{(Z-\delta\Psi)} = \frac{m^2_Z}{<\Psi>}$. Thus the inverse of $<\Psi>$ plays 
  a crucial role in determining the coupling strength
 between radion and SM fields. In the present case, we obtain
  \begin{equation}
   <\Psi> = \sqrt{\frac{12M^3}{k}}\exp{(-k\pi r_c)}\exp{[-\frac{l^2}{6}\exp{(-2u\pi r_c)}]}
   \nonumber\\
  \end{equation}
  Hence finally we arrive at,
  \begin{eqnarray}
   \lambda_{(H-\delta\Psi)}&=&\mu^2 \sqrt{\frac{12M^3}{k}}\exp{(k\pi r_c)}\nonumber\\
&\exp&{[\frac{l^2}{6}\exp{(-2u\pi r_c)}]}
   \label{coupling 1}
  \end{eqnarray}
  and
  \begin{eqnarray}
   \lambda_{(Z-\delta\Psi)}&=&m^2_Z \sqrt{\frac{12M^3}{k}}\exp{(k\pi r_c)}\nonumber\\
&\exp&{[\frac{l^2}{6}\exp{(-2u\pi r_c)}]}
   \label{coupling 2}
  \end{eqnarray}
  Once again we observe the appearance of an additional correction to the coupling from that
  obtained in \cite{kribs}. The correction, though small in general, can be significant for small
  values of $u r_c$ in the parameter space. These couplings however become same as that
  obtained in \cite{GW_radion} in the leading order of $u/k$. 
In summary, we obtain the stabilisation of a backreacted warped geometry model 
via the GW mechanism. In addition we study the radion phenomenology and obtain 
an equivalence to the results obtained in \cite{GW_radion, kribs}. 

\section*{Conclusion}\label{conclusion}
Stabilisation of RS braneword scenario is an essential requirement to study various implications 
of the presence of a warped extra dimension on particle phenomenology as well as  cosmology
and gravitational Physics. The stabilisation mechanism originally 
proposed by Goldberger and Wise considered a negligible backreaction 
on the background geometry.  In a more generalized version, the modified warp factor due to the influence of backreaction of a bulk scalar on the background geometry was derived in \cite{kribs} for certain class of scalar potentials with quartic self-interaction. \\
In this work we show that such a backreacted RS warped geometry model can also be stabilised
via the GW mechanism. The new stabilisation condition for the 
modulus is determined and its dependence on the backreaction parameter is found.
We also show that the radion phenomenology can be studied in this
background via the mechanism followed in \cite{GW_radion}. In the present work, the derived
mass and coupling of radion  have remarkable equivalence with the results 
obtained in \cite{kribs}, despite the difference in approaches between these two works.
Moreover  to find solutions to modified Einstein's equation and the  
scalar field equations in presence  of the backreaction of the scalar field,
a quartic form of scalar field potential have been considered in \cite{kribs}. However the GW stabilising scenario assumes only 
a quadratic mass term in the scalar potential  with negligible backreaction on the background geometry. 
Our result depicts a vital correlation between these two forms of potentials  in the leading order of $u$.
As a result the stabilisation condition, radion mass and  coupling parameter 
in these two different formalism get correlated  in the leading order of  parameter $u$. 
Our results also bring out how the backreaction effects modify the mass and coupling parameters of the model resulting into  modifications of the particle  phenomenology on the visible 3-brane.

\end{document}